\documentclass[a4paper,10pt,twoside]{cpc-hepnp}
\usepackage{upgreek,fancyhdr}
\usepackage{multicol,color}
\usepackage{graphicx,subfigure,dcolumn}
\usepackage{epsfig}
\usepackage{booktabs,slashed}
\usepackage{amssymb,bm,mathrsfs,bbm,amscd,amsfonts}
\usepackage[tbtags]{amsmath}
\usepackage{lastpage}
\usepackage{multirow}
\usepackage{amsmath}
\usepackage{cases}
\usepackage{tikz}
\usepackage[all,pdf]{xy}
\usepackage{graph icx}
\usepackage[colorlinks=true]{hyperref}

\begin{document}

\title{\boldmath Entanglement Entropy of Annulus in Holographic Thermalization
\thanks{We are very grateful to Li Li, Chao Niu, Qiang Wen and Cheng-Yong Zhang for helpful discussions and suggestions. This work is supported by the Natural Science Foundation of China
under Grant No. 11575195, 11875053 (Y.L.) and 11847229 (Z.-Y.X.). Y.L. also acknowledges the support from 555 talent project of Jiangxi Province. Z.-Y.X. also acknowledges the support from the National Postdoctoral Program for Innovative Talents BX20180318.}
}

 \author{Yi Ling     $^{1,2}$ \email{lingy@ihep.ac.cn}
 \quad Yuxuan Liu   $^{1,2}$ \email{liuyuxuan@ihep.ac.cn}
 \quad Zhuo-Yu Xian $^{3}$   \email{xianzy@itp.ac.cn}}

\maketitle

\address{
$^{1}$ Institute of High Energy Physics, Chinese Academy of Sciences, Beijing 100049, China\ \\
$^{2}$ School of Physics, University of Chinese Academy of Sciences,
Beijing 100049, China\ \\
$^{3}$Institute of Theoretical Physics, Chinese Academy of Science, Beijing 100190, China}

\maketitle

\begin{abstract}
The thermalization process of the holographic entanglement entropy
(HEE) of an annular domain is investigated over the Vaidya-AdS
geometry. We numerically determine the Hubeny-Rangamani-Takayanagi
(HRT) surface which may be a hemi-torus or two disks, depending on
the ratio of the inner radius to the outer radius of the annulus.
More importantly, for some fixed ratio of two radii, it
undergoes a phase transition or double phase transitions from a
hemi-torus configuration to a two-disk configuration, or vice
versa, during the thermalization. The occurrence of various
phase transitions is determined by the ratio of two radii of the
annulus. The rate of entanglement growth is also investigated
during the thermal quench. The local maximal rate of entanglement
growth occurs in the region with double phase transitions.
Finally, if the quench process is fairly slow which may be
controlled by the thickness of null shell, the region with double
phase transitions vanishes.
\end{abstract}

\begin{keyword}
  holographic entanglement entropy, thermal quench, annulus
  \end{keyword}

  \begin{pacs}
    11.25Tq
    \end{pacs}

\section{Introduction}
Entanglement entropy, as a vital tool to measure the entanglement between quantum systems, has been extensively investigated in recent years. For a strongly coupled quantum system which is in a pure state, the entanglement entropy between the subsystem $\mathcal{A}$ and its complement $\bar{\mathcal{A}}$ is proportional to the area of the boundary $\partial \mathcal{A}$ to the leading order \cite{Rangamani:2016dms}. In the context of AdS/CFT correspondence \cite{Maldacena:1997re,Gubser:1998bc,Witten:1998qj}, the Ryu-Takayanagi (RT) formula \cite{Ryu:2006bv,Ryu:2006ef} conjectures that the entanglement entropy can be evaluated by the area of the minimal surface $\gamma_{\mathcal{A}}$ in the bulk which is homologous to the subregion $\mathcal{A}$ on the boundary. Such a surface is also called RT surface or its covariant version, HRT surface \cite{Hubeny:2007xt}. In literature, the RT formula has been extensively testified in
various holographic models and has been specifically computed for
the subregion with a variety of shapes \cite{Fonda:2014cca,Fonda:2015nma}. In particular, its significant role in diagnosing the quantum critical phenomena in strongly coupled systems has been disclosed in \cite{Holzhey:1994we,Vidal:2002rm,Ling:2016dck,Kitaev:2005dm,Calabrese:2004eu,Grover:2011fa,Ling:2016wyr,Ling:2015dma,Guo:2019vni,Zeng:2015tfj,Zeng:2015wtt,Zeng:2016sei}.

Investigating the dynamical behavior of the response such as entanglement entropy by perturbing the system away from the equilibrium state is very crucial for characterizing the feature of a non-equilibrium system. A simple case is the evolution of a system after a quench process, which can be realized by turning on an external source for a short time. As a result, the system will be excited and subsequently equilibrates under evolution. In the holographic duality, the thermal quench can be modeled by Vaidya-AdS geometry, which describes the collapsing process of a null shell, which initially falls from the boundary of AdS to the bulk and eventually forms a Schwarzschild-AdS (SAdS) black brane.

In literature, the evolution behavior of HEE during such a quench
process has previously been studied in
\cite{Albash:2010mv,Balasubramanian:2010ce,Balasubramanian:2013rva,Liu:2013qca,Caceres:2012em,Liu:2013iza,Camilo:2014npa,Chen:2018mcc,Ling:2018xpc,Zhou:2019jlh,Bai:2014tla,Li:2013sia},
for a subregion with the shape of strip and disk. For an annular
subsystem $\mathcal{A}$, the research on its thermalization
process is still lacking. Previously, the HEE for such subregion
has been computed for a static background
\cite{Han:2019scu,Nakaguchi:2014pha}. It is interesting to notice
that there exist two possible configurations for the HRT surface.
One has the hemi-torus shape \cite{Drukker:2005cu,Dekel:2013kwa}
while the other has the two-disk shape
\cite{Shiba:2010dy,Shiba:2012np,Cardy:2013nua}. Which
configuration the HRT surface takes depends on the ratio of the
inner radius to the outer radius of the annulus. Moreover, the
rate of entanglement growth is captured by the ``entanglement
tsunami'' diagram \cite{Liu:2013iza,Liu:2013qca}, which treats
the null shell as an entanglement wave entangling the region
$\mathcal{A}$ with the outside. Since the growth rate generally
depends on the ratio of two radii, it is intriguing to investigate
the phase transition of the HRT surface $\gamma_\mathcal{A}$ as
well as the maximal rate of entanglement growth during the
holographic quench process.

The paper is organized as follows. In Sec.2, we introduce the
setup for the Vaidya-AdS background. The integral expressions for
the area of the HRT surface will be derived for a subregion with
the shape of annulus, and in the pure AdS background, the mutual
information across the annular subsystem $\mathcal{A}$ will be
briefly discussed. In Sec.3, we present our numerical results for
the time dependence of the HEE during the quench. In addition,
various phase transitions will be illustrated in detail. The
maximal rate of entanglement growth in each case and the
dependence on the thickness of the shell will also be discussed.
Sec.4 is our conclusion and discussions.

\begin{figure}
  \centering
\subfigure[]{\label{fig_ht1}
\includegraphics[width=150pt]{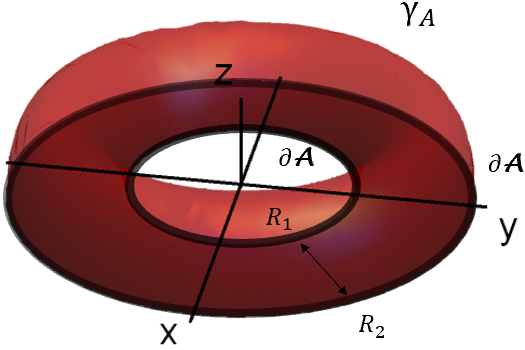}}
\hspace{30pt}
\subfigure[]{\label{fig_ht2}
\includegraphics[width=150pt]{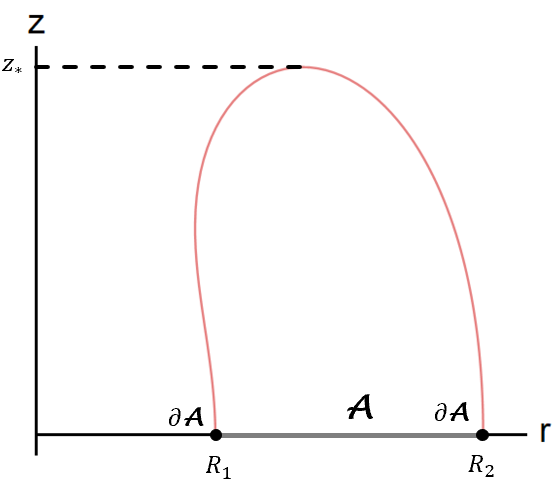}}
\hspace{80pt}
\subfigure[]{\label{fig_2s1}
\includegraphics[width=150pt]{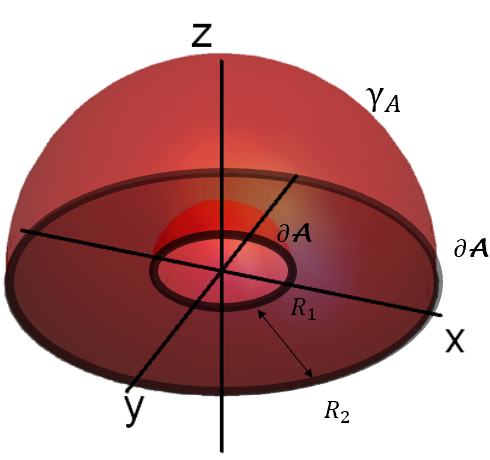}}
\hspace{30pt}
\subfigure[]{\label{fig_2s2}
\includegraphics[width=150pt]{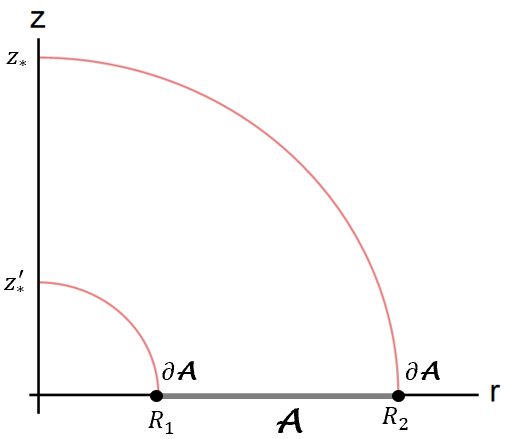}}

\caption{Fig.\ref{fig_ht1} illustrates the extremal surface $\gamma_\mathcal{A}$ in the hemi-torus shape. Fig.\ref{fig_ht2} is its cross section after suppressing the $\theta$ direction. While Fig. \ref{fig_2s1} illustrates the extremal surface $\gamma_\mathcal{A}$ in the two-disk shape and Fig.\ref{fig_2s2} is its cross section after suppressing the $\theta$ direction.}\label{fig_transition}
\end{figure}

\section{The Setup}
In this section, we firstly introduce the Vaidya-AdS$_4$ metric, which describes the geometry of collapsing a null shell falling from the boundary to form a SAdS black hole. Then we will derive the expression of the area functional of an annular domain $\mathcal{A}$. For intuition, we will demonstrate two possible configurations of HRT surface $\gamma_\mathcal{A}$ in the pure AdS$_4$ case. Finally, we will discuss the characteristics of mutual information across the annulus.

\subsection{Vaidya-AdS$_4$ background}
Consider the Vaidya-AdS$_4$ metric in the
Eddington-Finkelstein coordinates
\begin{equation}\label{eq_vaidya-ads}
  ds^2=\frac{1}{z^2}\left(-f(v,z)dv^2-2dvdz+dr^2+r^2d\theta^2\right),
\end{equation}
with
\begin{equation}   f(v,z)=1-\frac{M}{2}\left(1+\tanh
\frac{v}{v_{0}}\right)z^3,\nonumber
\end{equation}
where we have set the AdS radius $R_{AdS}=1$. $M$ characterizes the mass of the black hole and $v_0$ labels the thickness of the null shell. In this setup, the coordinate $v$ labels the boundary time $t$ when $z\rightarrow 0$. Moreover, in the limit of $v\rightarrow -\infty$, the metric in (\ref{eq_vaidya-ads}) approaches $$f(v,z)=1,$$ which is the AdS metric, while in the limit of $v\rightarrow \infty$, the metric approaches $$f(v,z)=1-M z^3,$$ which is just the metric on the SAdS spacetime.

\begin{figure}
  \centering
\subfigure[]{\label{fig_mi1}
\includegraphics[width=150pt]{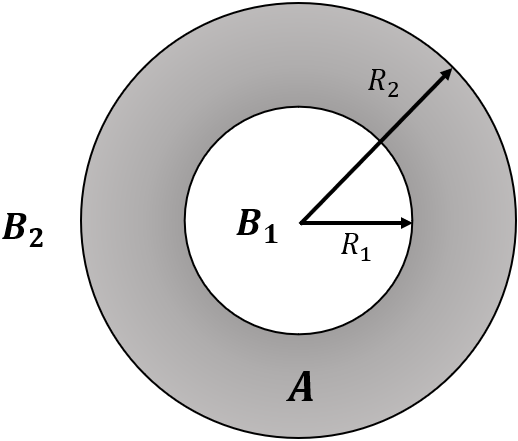}}
\hspace{30pt}
\subfigure[]{\label{fig_mi2}
\includegraphics[width=180pt]{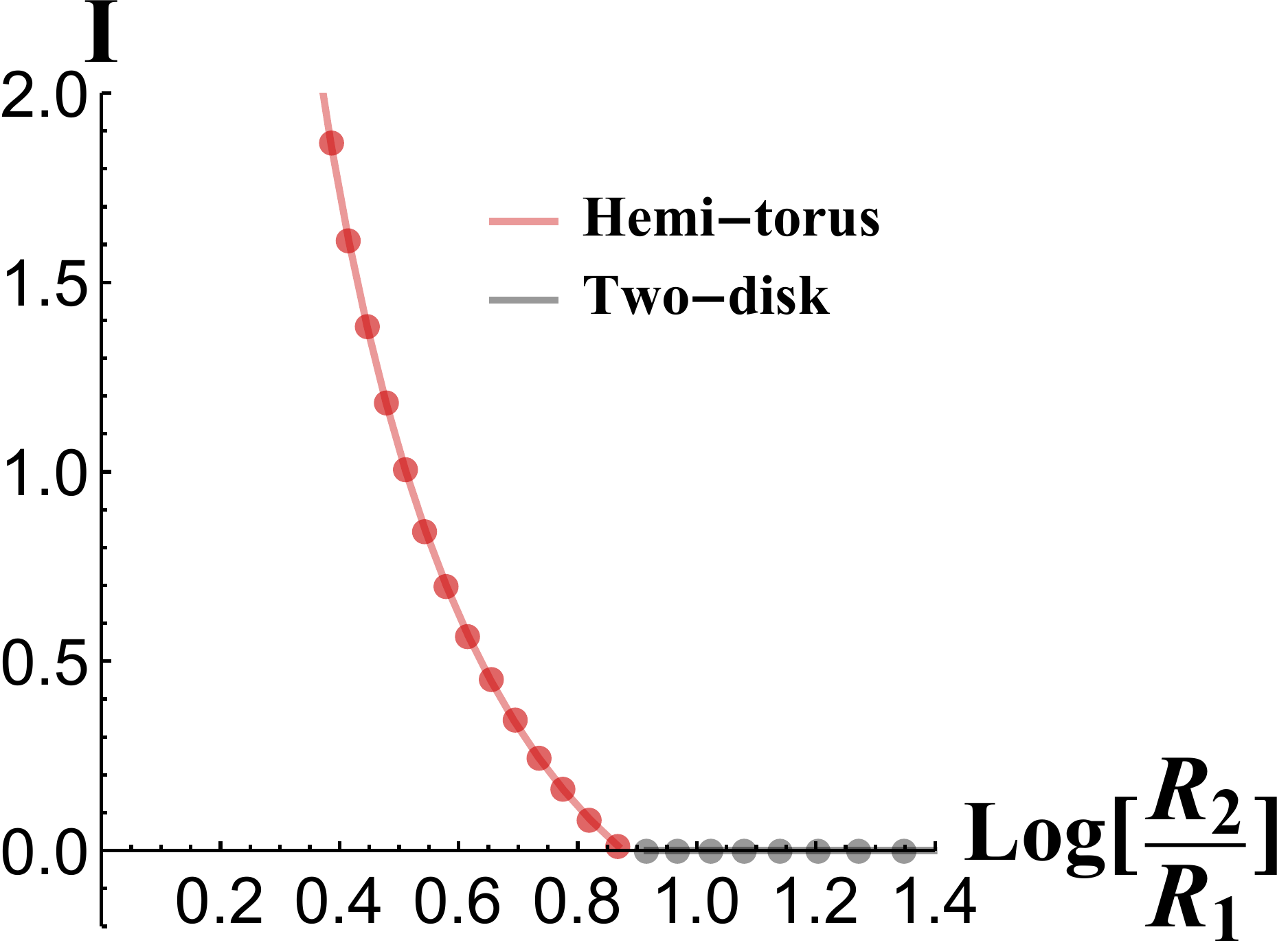}}

\caption{As shown in Fig.\ref{fig_mi1}, $\mathcal{A}$ represents the subsystem we investigate, while $\mathcal{B}_1$ and $\mathcal{B}_2$ are two disjoint subsystems separated by $\mathcal{A}$. In Fig.\ref{fig_mi2}, the holographic mutual information $I$ between $\mathcal{B}_1$ and $\mathcal{B}_2$ is plotted with different ratio of two radii $R_1$ and $R_2$ in the AdS$_4$ background. The phase transition occurs at $R_2/R_1\approx2.4$.}\label{fig_mi}
\end{figure}

\subsection{HRT surface $\gamma_{\mathcal{A}}$ of an annular domain $\mathcal{A}$}
On the boundary, consider the subregion $\mathcal{A}(r,\theta)$ which is an annulus with $r\in\left[R_1,R_2\right]$ and $\theta\in\left(0,2\pi\right]$. Due to the spherical symmetry, the region $\mathcal{A}$ is completely specified by the radius $r$. Then the corresponding area of the extremal surface $\gamma_\mathcal{A}$ anchored on $\partial\mathcal{A}$ is described by
$$z=z(r), \qquad v=v(r)$$
 and reads as
\begin{equation}\label{eq_vaidya-HRT}
  A[\gamma_\mathcal{A}]=2\pi\int_{R_1}^{R_2}dr\frac{r}{z^{2}}\sqrt{1-2v'z'-f(v,z)v'^2}.
\end{equation}
The equations of motion are obtained by extremizing the area functional (\ref{eq_vaidya-HRT}). It is noticed that as the inner radius $R_1\rightarrow0$, the above area functional reduces to the functional corresponding to a spherical region $\mathcal{A}$ with radius $R_2$ (see \cite{Albash:2010mv,Liu:2013qca,Camilo:2014npa}).

Before the thermal quench, the geometry is a pure AdS$_4$ spacetime and the corresponding area functional reduces to
\begin{equation}\label{eq_ads-HRT}
  A[\gamma_\mathcal{A}]=2\pi\int_{R_1}^{R_2}dr\frac{r}{z^{2}}\sqrt{1+z'^2}.
\end{equation}
The phase transition of HRT surface in AdS$_4$ spacetime has been investigated in literature. As the ratio of the outer radius $R_2$ to the inner radius $R_1$ approaches one, the HRT surface is in the hemi-torus phase (Fig.\ref{fig_ht1}). Then decreasing the inner radius $R_1$, as the ratio $R_2/R_1\rightarrow 2.4$ \cite{Nakaguchi:2014pha}, the configuration of the HRT surface will undergo a phase transition from the hemi-torus phase to the two-disk phase (Fig.\ref{fig_2s1}).

\begin{figure}
  \centering
\subfigure[]{\label{fig_annang}
\includegraphics[width=150pt]{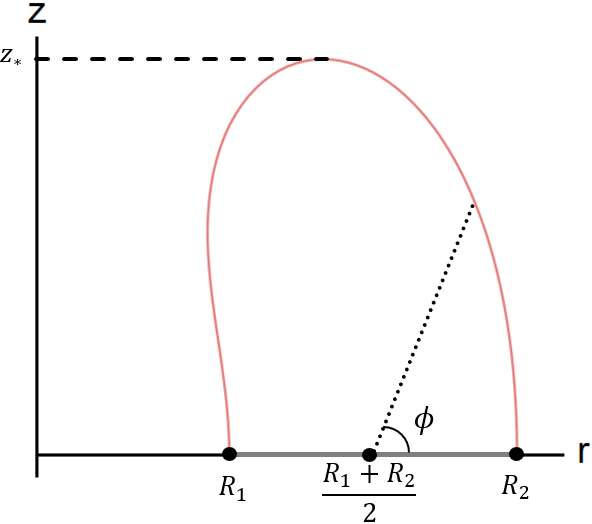}}
\hspace{30pt}
\subfigure[]{\label{fig_sphang}
\includegraphics[width=150pt]{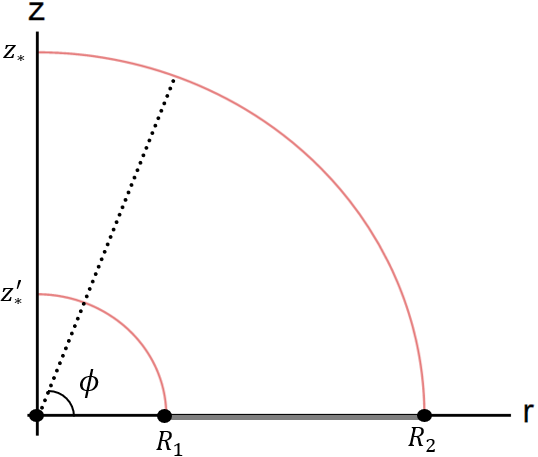}}

\caption{The parameterization of the HRT surface. If the HRT
surface is in the hemi-torus phase, the parameter $\phi$ is
shown in Fig.\ref{fig_annang} with $0\leq \phi \leq \pi$. While
if the HRT surface is in the two-disk phase, the parameter is
shown in Fig.\ref{fig_sphang} with $0 \leq \phi \leq
\frac{\pi}{2}$.}\label{fig_angle}
\end{figure}

\subsection{Mutual information}
Mutual information between two disjoint subsystems $B_1$ and $B_2$ is defined as
\begin{equation}
  I(B_1 ; B_2) \equiv S(B_1)+S(B_2)-S(B_1 \cup B_2) \geq 0
\end{equation}
Specifically, for an annular subregion $\mathcal{A}$ of the boundary, we take $\mathcal{B}_1$ and $\mathcal{B}_2$ as two disjoints subsystems, which are located in $r\leq R_1$ and $r\geq R_2$ respectively as shown in Fig.\ref{fig_mi1}.

In the holographic setup, when decreasing the inner radius $R_1$ with fixed $R_2$, the mutual information between $\mathcal{B}_1$ and $\mathcal{B}_2$ decreases monotonically to zero, which is consistent with the result in literature \cite{Fonda:2014cca,Han:2019scu,Nakaguchi:2014pha}. Furthermore, the HRT surface $\gamma_\mathcal{A}$ corresponding to the annulus $\mathcal{A}$ is in the hemi-torus phase as $I(\mathcal{B}_1;\mathcal{B}_2)>0$, while it is in the two-disk phase as $I(\mathcal{B}_1;\mathcal{B}_2)=0$ as shown in Fig.\ref{fig_mi2}.

For a general quantum system, mutual information measures the
entanglement and correlations between the subsystems and gives an
upper bound for the correlations as well. Therefore, for a system
in pure state, if a HRT surface $\mathcal{A}$ is in the hemi-torus
phase, the d.o.f. of the subsystem $\mathcal{B}_1$ are generally
entangled with those of the subsystem $\mathcal{B}_2$. While, if
the HRT surface $\gamma_\mathcal{A}$ is in the two-disk phase,
there is no entanglement between them.

So far, we have derived the integral expressions of the
entanglement entropy between the subsystem $\mathcal{A}$ and its
complement and discussed the mutual information across the
subsystem $\mathcal{A}$ in pure AdS$_4$ spacetime. In the next
section, we will investigate the evolution of the HEE in the
Vaidya-AdS$_4$ spacetime.

\section{Numeric Method}
First of all, to get rid of ultra-violet (UV) divergence at $z\rightarrow0$, we only consider the finite term in (\ref{eq_vaidya-HRT}), which is
\begin{align}
  A_{Ren}=A[\gamma_\mathcal{A}]-\frac{R_1+R_2}{\epsilon},
\end{align}
where $\epsilon$ is the UV cut-off. It is manifest that $A_{Ren}$ is cut-off independent. Then we fix all the free parameters: the mass $M=1$, the thickness of the shell $v_0=0.3$, the outer radius of the annulus $R_2=5$ and the inner radius $R_1\in[0.5,4.5]$.

Specifically, the extremal surface $\gamma_{\mathcal{A}}$ is parameterized by
\begin{equation}\label{eq_para}
  z=z(\phi), \quad r=r(\phi), \quad v=v(\phi),
\end{equation}
where $\phi$ is the polar angle as shown in Fig.\ref{fig_angle}.
Since above three variables in (\ref{eq_para}) are not
independent, it is necessary to introduce a constraint equation.
In the hemi-torus phase, the constraint equation is
\begin{equation}\label{eq_ceqann}
  z(\phi)\cos(\phi)-\left(r(\phi)-\frac{R_1+R_2}{2}\right)\sin(\phi)=0, \qquad (0\leq\phi\leq\pi)
\end{equation}
 and the boundary conditions are $$z(0)=z(\pi)=0, \quad v(0)=v(\pi)=t, \quad r(0)=R_2, \quad r(\pi)=R_1.$$ While in the two-disk phase, the constraint equation is
\begin{equation}\label{eq_ceqsph}
  z(\phi)\cos(\phi)-r(\phi)\sin(\phi)=0,\qquad (0\leq\phi\leq\frac{\pi}{2})
\end{equation}
 and the boundary conditions reduce to $$z(0)=z'(\frac{\pi}{2})=0, \quad v(0)=t, \quad v'(\frac{\pi}{2})=0,\quad r(0)=R_2\;(\text{or} \; R_1), \quad r(\frac{\pi}{2})=0. $$ The above constraint equations (\ref{eq_ceqann}) and (\ref{eq_ceqsph}) are imposed to the area functional (\ref{eq_vaidya-HRT}) by the method of lagrange multiplier. Then, the corresponding E.O.M. can be numerically solved by the method of finite differences.

 \begin{figure}
  \centering
  \subfigure[]{\label{fig_listR1}
\includegraphics[height=150pt]{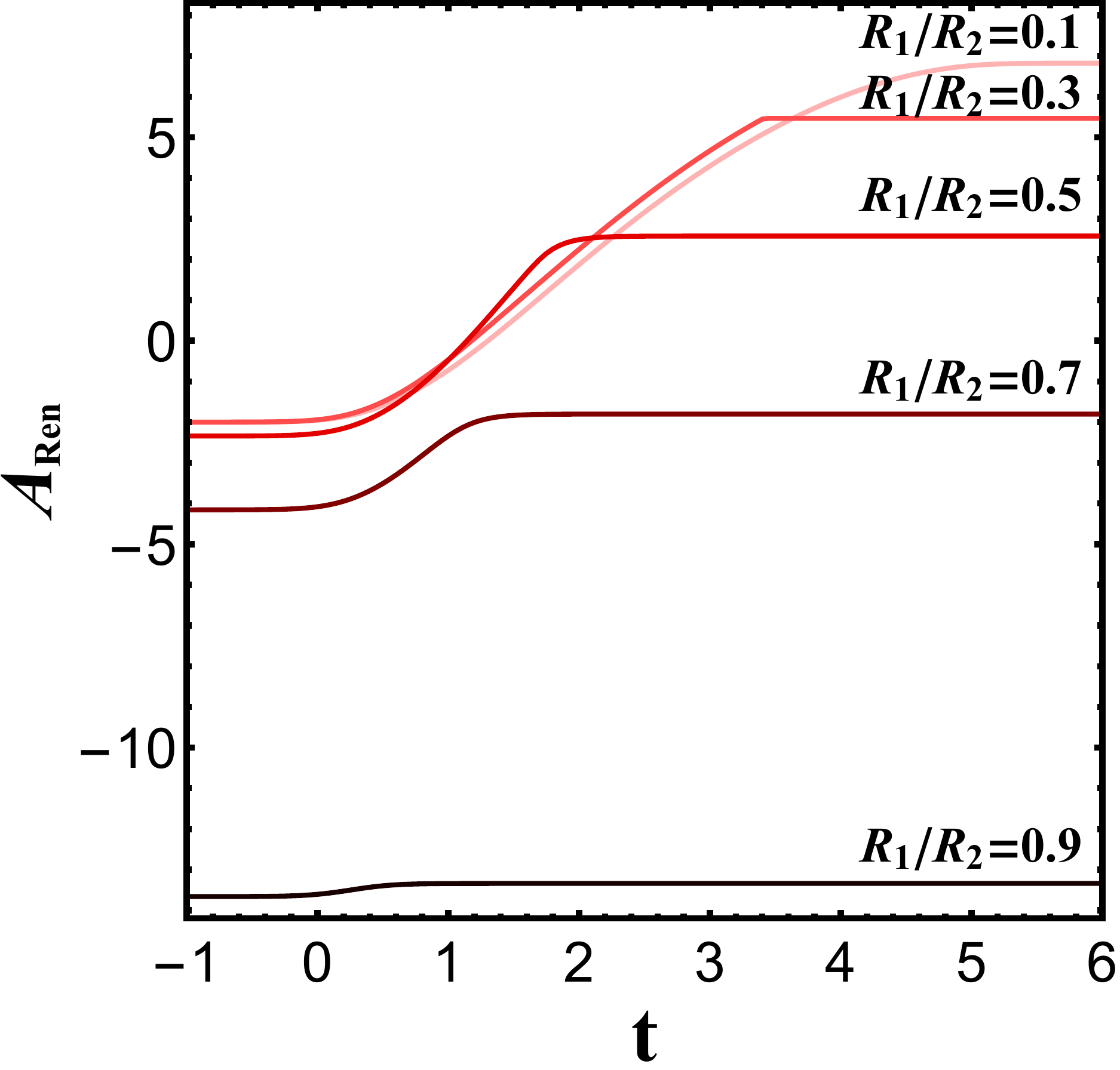}}
\hspace{0pt}
\subfigure[]{\label{fig_sot}
\includegraphics[height=150pt]{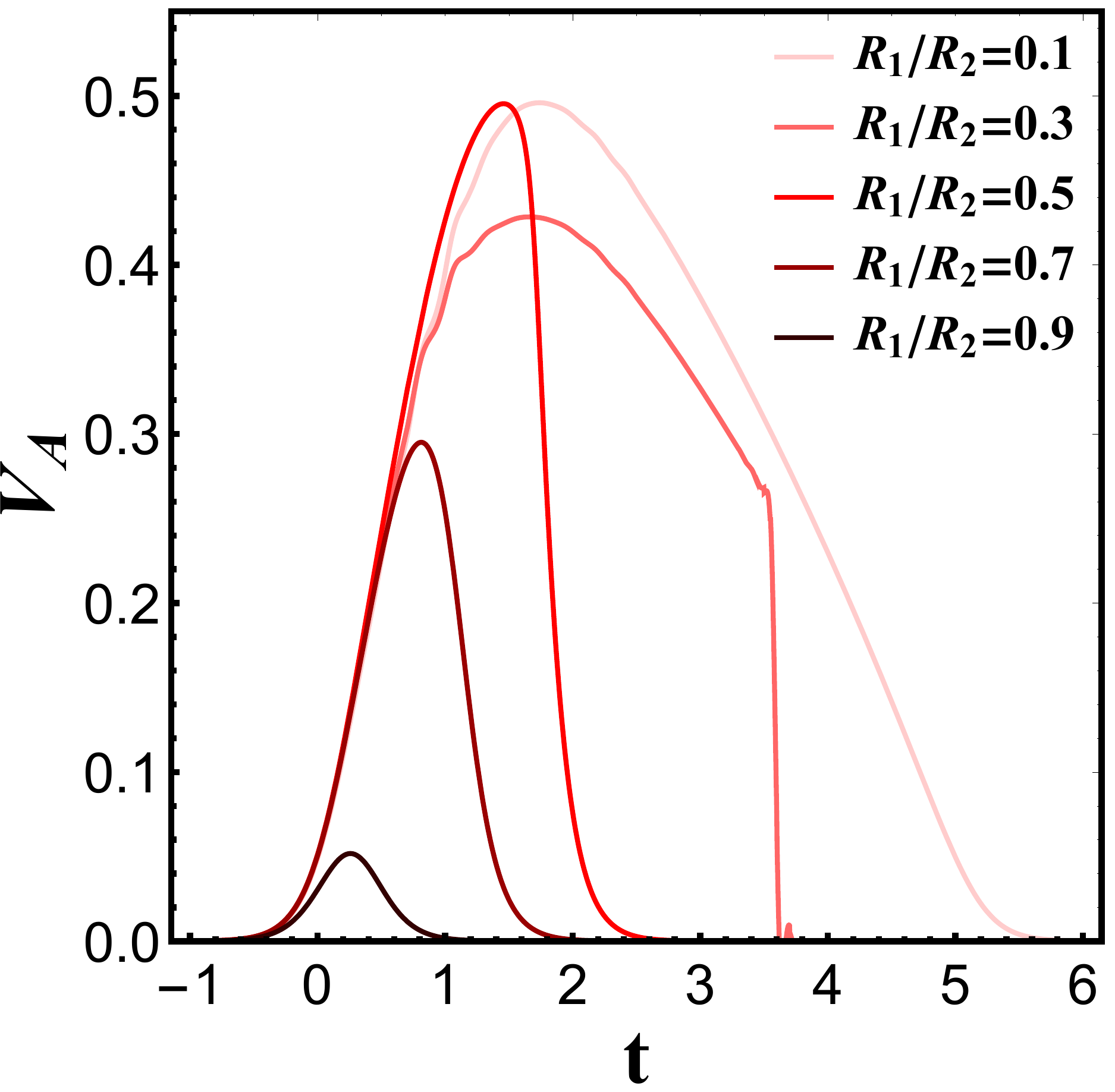}}
\hspace{0pt}
\subfigure[]{\label{fig_contour}
\includegraphics[height=150pt]{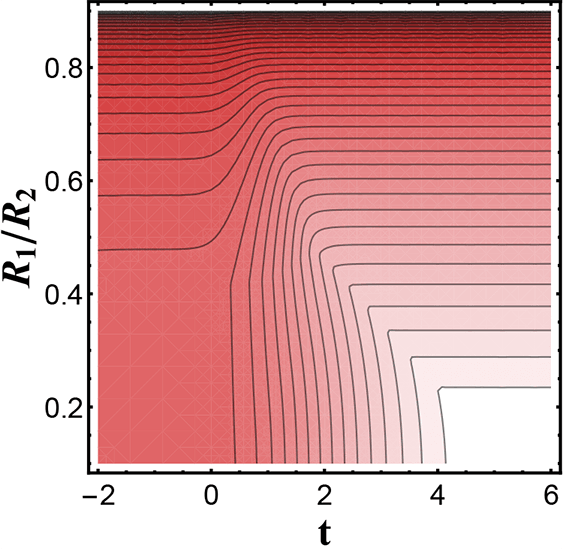}}

\caption{Fig.\ref{fig_listR1} illustrates the evolution of the HEE
with $R_1/R_2=0.1, 0.3, 0.5, 0.7$ and $0.9$, respectively, while
Fig.\ref{fig_sot} illustrates the rate of the entanglement growth
in the unit of the length of $\partial\mathcal{A}$.
Fig.\ref{fig_contour} is the contour plotting of HEE during the
evolution with different ratio $R_1/R_2$. The lighter the color
is, the larger the value of HEE is.}
\end{figure}

 \begin{figure}
  \centering
\subfigure[]{\label{fig_transition03}
\includegraphics[width=150pt]{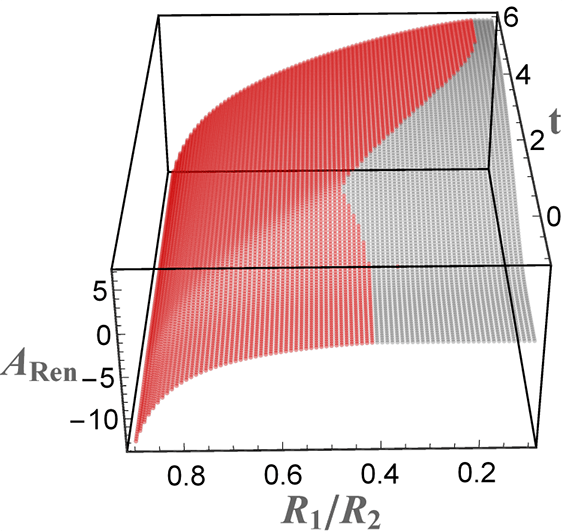}}
\hspace{30pt}
\subfigure[]{\label{fig_transition032d}
\includegraphics[width=150pt]{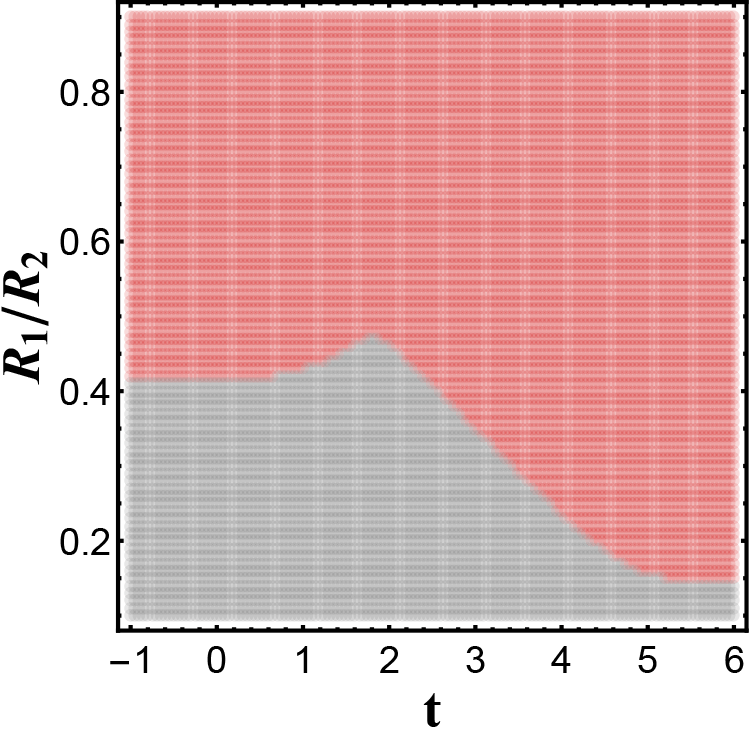}}

\caption{The evolution of HEE with different $R_1/R_2$ is shown in Fig.\ref{fig_transition03}. The data in red indicate that the HRT surface is in the hemi-torus phase, while the data in grey indicate that the HRT surface is in the two-disk phase. Fig.\ref{fig_transition032d} is obtained by projecting the result in Fig.\ref{fig_transition03} onto the $(R_1/R_2,t)$ plane, which is convenient for us to identify the phase corresponding to the parameter $R_1/R_2$ at any moment. }\label{fig_transition033}
\end{figure}

\begin{figure}
  \centering
\includegraphics[width=250pt]{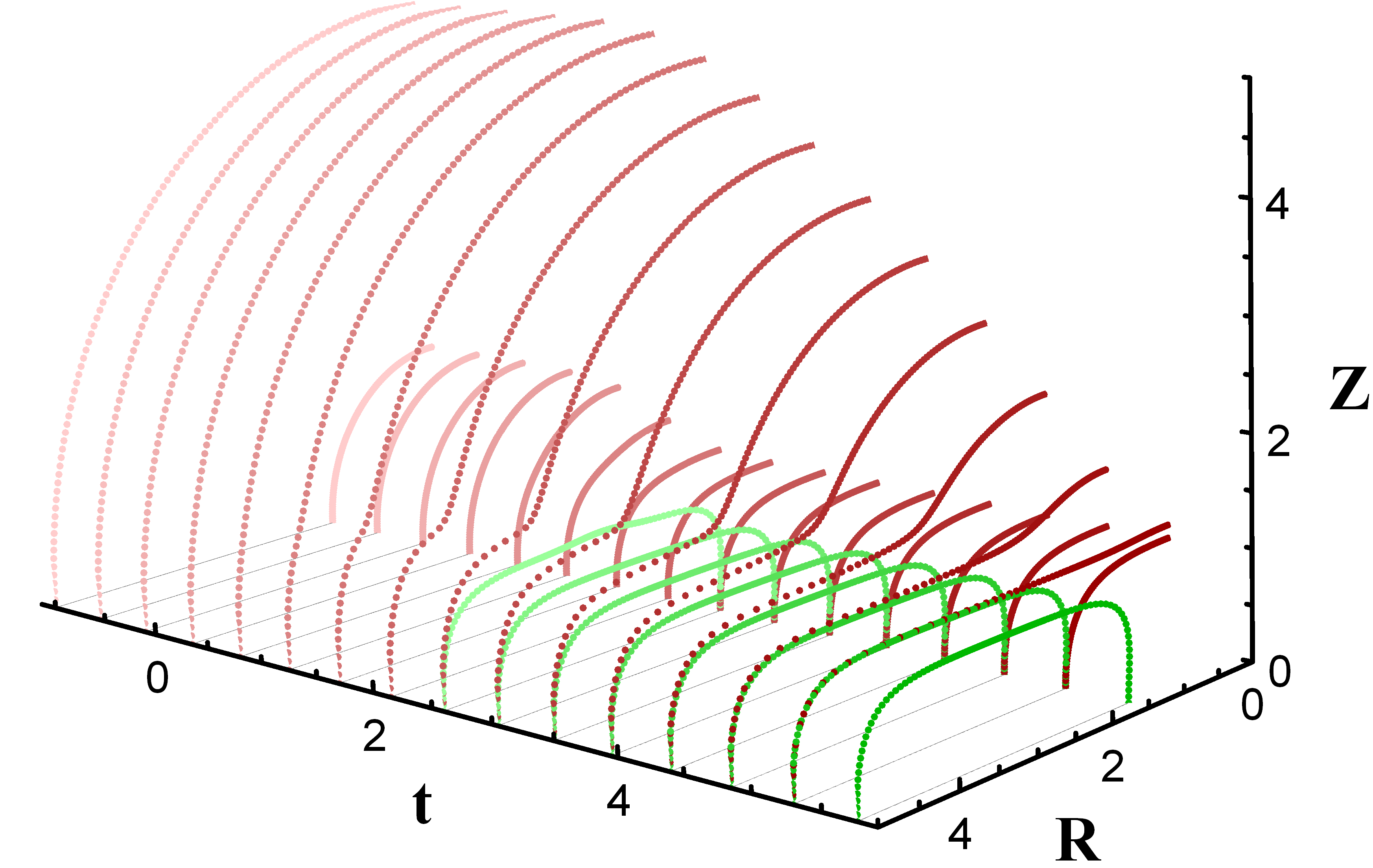}

\caption{Two candidates of the HRT surface anchored on the
$\partial\mathcal{A}$ are shown in the figure. At each moment
during the evolution, the candidate in the two-disk phase is
plotted in red, while the candidate in the hemi-torus phase is
plotted in green. Note that the HRT surface will not
exactly lie on a time slice during the evolution, but the
figure is good enough for us to gain
intuition.}\label{fig_candidates}
\end{figure}

Next we turn to present the numeric results for the evolution of
the HEE based on renormalized area $A_{Ren}$. Firstly, we are
concerned with the time dependence of the HEE for different ratio
of two radii $R_1/R_2$, as illustrated in Fig.\ref{fig_listR1}. It
is obvious to find that the entanglement entropy is always
increasing roughly with a linear manner\footnote{The
linearity is not quite precise here for small $R_1/R_2$ as
shown in Fig.\ref{fig_sot}, but in general, the larger the size of
the region $\mathcal{A}$ is, the more obvious the stage of linear
growth is.} with time at the intermediate stage of the thermal
quench and finally saturates. Moreover, defining the rate of
entanglement growth as $$V_\mathcal{A}=\frac{1}{R_1+R_2}\frac{d A_{Ren}}{dt},$$ we find it always increases with time
at the early stage and eventually decreases to zero at equilibrium
as shown in Fig.\ref{fig_sot}. In general, the saturation value as
well as the saturation time is increasing with the width of the
region $\mathcal{A}$, which are quite common phenomena in
literature. Because the HRT surface with the wider boundary region
$\mathcal{A}$ usually stretches deeper into the bulk region, the
null shell also takes longer time to reach this region during the
holographic quench process. As a consequence, it takes longer time
to get saturation. Furthermore, the saturation time approaches a
constant as the ratio $R_1/R_2\rightarrow0$ as shown in
Fig.\ref{fig_contour}. This result indicates that in the region
where $R_1/R_2$ approaches zero, the HRT surfaces with the
different $R_1$ are in the two-disk phase near saturation. Since
the outer radius $R_2$ is fixed in the setting, all the HRT
surfaces which are in the two-disk phase share the same outer part
of extremal surface as well as the saturation time.

It is noticed that the evolution of HEE undergoes an unsmooth
saturation while $R_1/R_2=0.3$. This result reveals that the HRT
surface undergoes a phase transition during the thermal quench.
Meanwhile, other results in Fig.\ref{fig_listR1} and \ref{fig_sot}
demonstrate that the HRT surfaces always in the same phase during
the quench. We will analyze these results with more details in the
next subsection.

\subsection{Phase transition of the HRT surface}\label{sec_3}

In Fig.\ref{fig_transition03} we demonstrate the time evolution of HEE under different inner radius $R_1$. The region marked in red represents the HRT surface in the hemi-torus phase, while the region in gray represents the HRT surface in the two-disk phase. In general, the HRT surfaces are in the hemi-torus phase as the ratio of two radii $R_1/R_2$ approaches one, otherwise, if  $R_1/R_2\ll1$, the HRT surfaces will be in the two-disk phase. In addition, the critical $R_1$, which may be defined as the borderline of two phases, shifts non-monotonically during the thermal quench as shown in Fig.\ref{fig_transition032d}. At the early stage of the quench, the critical point shifts towards the outer radius $R_2$. After reaching its peak at $t\approx1.80$, the critical point decreases monotonically to a lower level and eventually becomes stable.

The lower level of the critical point at late time can be
understood from Fig.\ref{fig_candidates}. At the early stage, the
candidate in the two-disk phase possesses a smaller area and
thus is the HRT surface in consequence. As the evolution
proceeds, the other candidate in the hemi-torus phase begins
to compete with the candidate in the two-disk phase and eventually
becomes the HRT surface at the late stage. Note that at the late
stage, the candidate in the two-disk phase possesses a thin
bottleneck near $R=0$ and this will naturally lead to the
candidate in hemi-torus phase becoming the HRT surface. In
addition, the larger the ratio $R_1/R_2$ is, the earlier the thin
bottleneck occurs. As a consequence, the phase transition occurs
earlier during the evolution.

It is interesting to notice that for a fixed ratio $R_1/R_2$, the phase transition occurs during the thermal quench. Further, the times of phase transition depend on the specific value of the ratio. According to this, the evolution of HEE during the quench process can be characterized by the following three distinct types.
\clearpage
\begin{figure}
  \centering
  \subfigure[]{\label{fig_transitionno1}
\includegraphics[height=150pt]{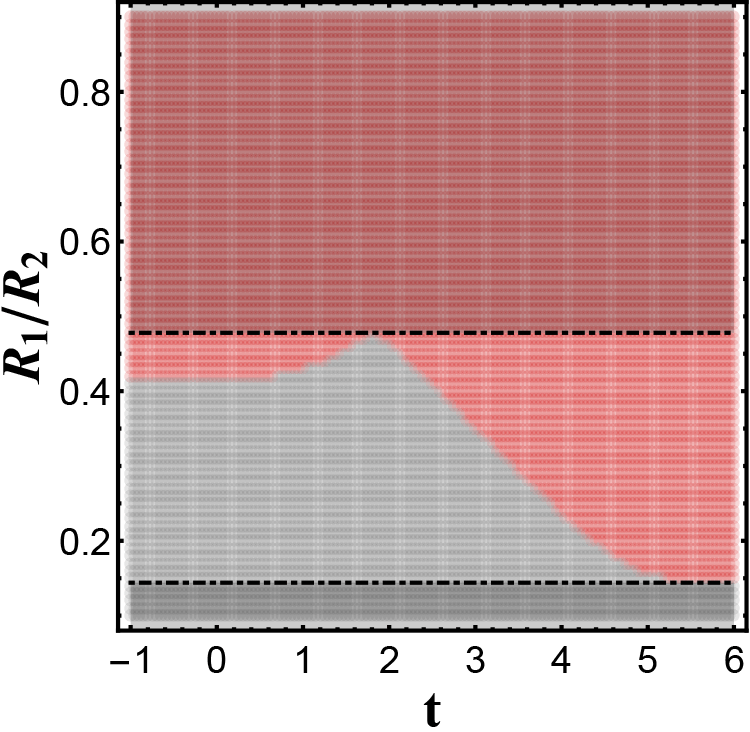}}
\hspace{0pt}
\subfigure[]{\label{fig_transitionsingle1}
\includegraphics[height=150pt]{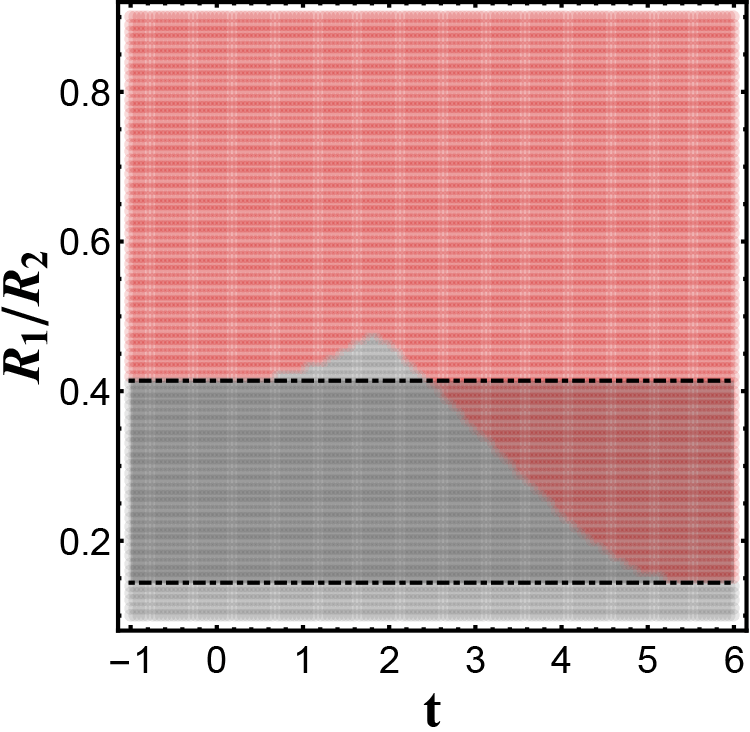}}
\hspace{0pt}
\subfigure[]{\label{fig_transitiondouble1}
\includegraphics[height=150pt]{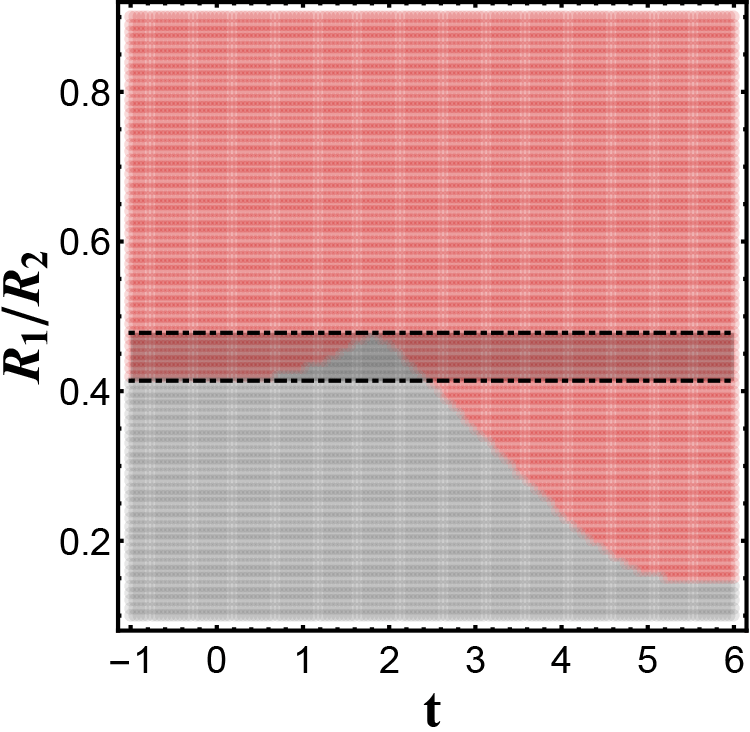}}
\hspace{0pt}
\subfigure[]{\label{fig_transitionno2}
\includegraphics[height=150pt]{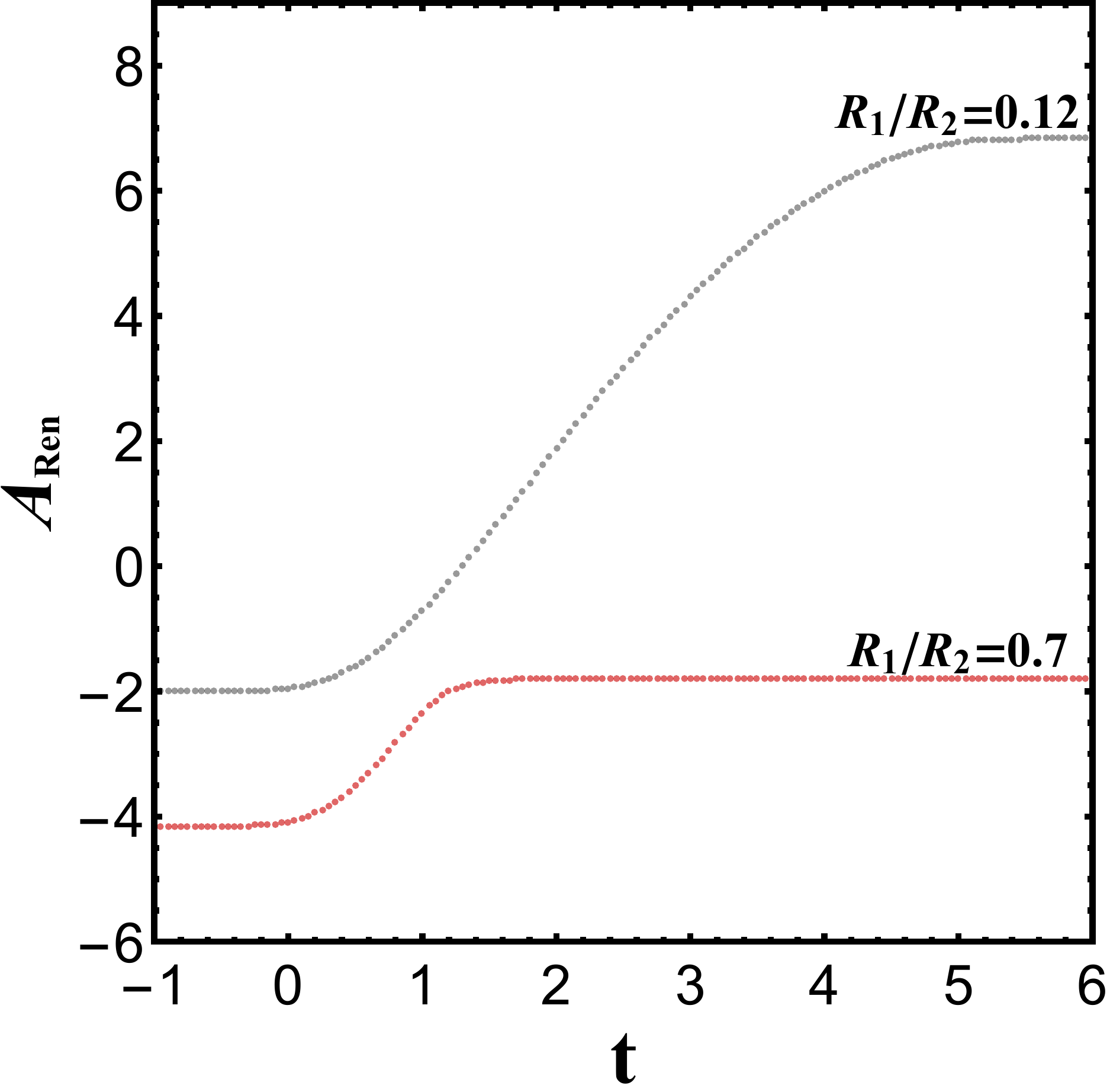}}
\hspace{0pt}
\subfigure[]{\label{fig_transitionsingle2}
\includegraphics[height=150pt]{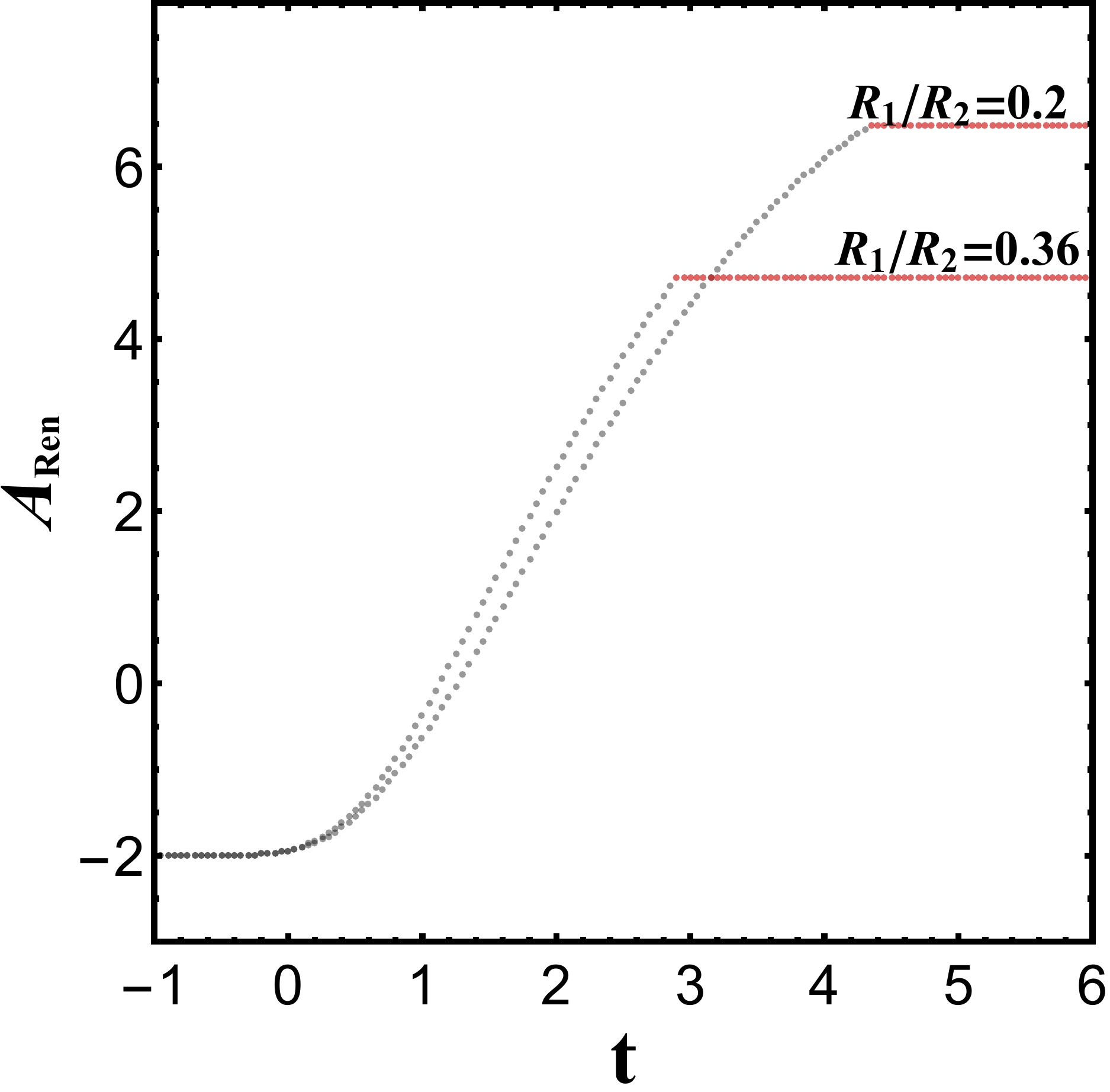}}
\hspace{0pt}
\subfigure[]{\label{fig_transitiondouble2}
\includegraphics[height=150pt]{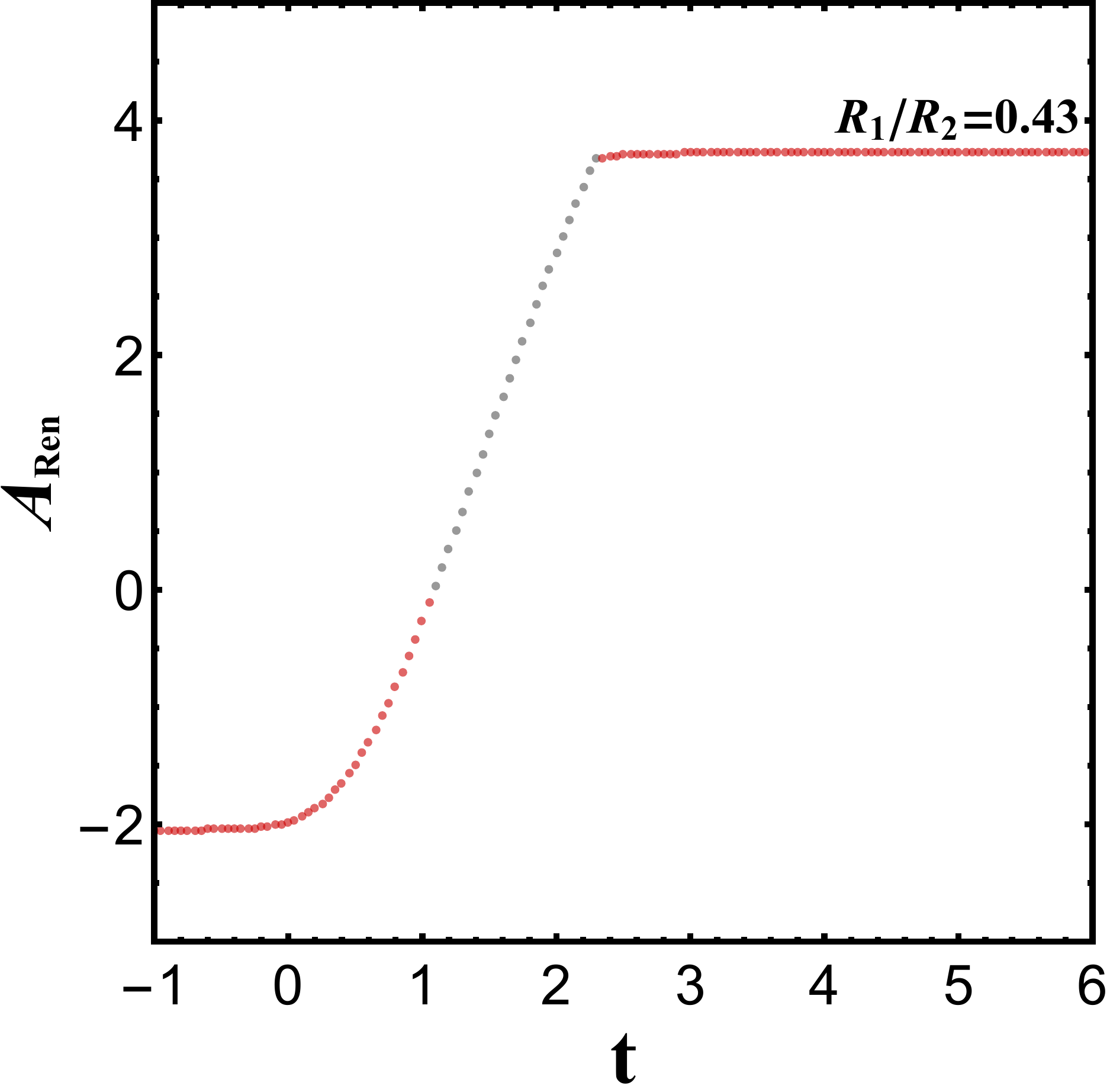}}

\caption{The shadow regions in Fig.\ref{fig_transitionno1},
Fig.\ref{fig_transitionsingle1} and
Fig.\ref{fig_transitiondouble1} represent the region with no phase
transition, single phase transition and double phase transitions
respectively, during the evolution of HEE with fixed $R_1/R_2$.
The curves in grey exhibit the evolution of HEE in the two-disk
phase while the curves in red exhibit the evolution of HEE in the
hemi-torus phase.}
\end{figure}

\begin{itemize}
\item Region with no phase transition

For the ratio $R_1/R_2\geq0.48$, the HRT surface is always in the
hemi-torus phase during the quench process, while for the ratio
$R_1/R_2\leq0.14$, the HRT surface is always in the two-disk phase
as shown in Fig.\ref{fig_transitionno1}. Note that without phase
transition, the evolution of HEE of an annular domain is similar
to that of a ball-shaped domain
\cite{Albash:2010mv,Caceres:2012em}, as the evolution curve is
always smooth (Fig.\ref{fig_transitionno2}). Moreover, the
subsystem $\mathcal{A}$ with greater $R_2-R_1$ generally possesses
a greater saturation value and longer saturation time as discussed
before.

\item Region with single phase transition

For the ratio $0.14 \leq R_1/R_2 \leq 0.42$, the HRT surface is in the two-disk phase at the early stage of evolution, then it will undergo a phase transition to the hemi-torus phase and persist all the way to saturation as shown in Fig.\ref{fig_transitionsingle1}. In addition, the critical point decreases almost linearly with time $t$ and ultimately reaches a global minimum which is consistent with the critical point in the Schwarzschild-AdS geometry.

When the system approaches the critical point, the first derivative of the HEE with respect to time $t$ is discontinuous (Fig.\ref{fig_transitionsingle2}). Similarly, in this region both the saturation value and the saturation time are generally increasing with $R_2-R_1$. During the evolution, the discontinuity of the derivative of HEE with respect to time is discussed mostly when the boundary region $\mathcal{A}$ is a strip and the width of the strip is greater than the event horizon. The difference is that when the boundary $\mathcal{A}$ is a strip, the discontinuity occurs due to the multiple values of the extremal surface \cite{Albash:2010mv,Caceres:2012em,Chen:2018mcc,Ling:2018xpc}, but when the boundary region $\mathcal{A}$ is an annulus, the discontinuity occurs due to the phase transition from the two-disk phase to the hemi-torus phase.

\clearpage

\begin{figure}
  \centering
  \includegraphics[width=275pt]{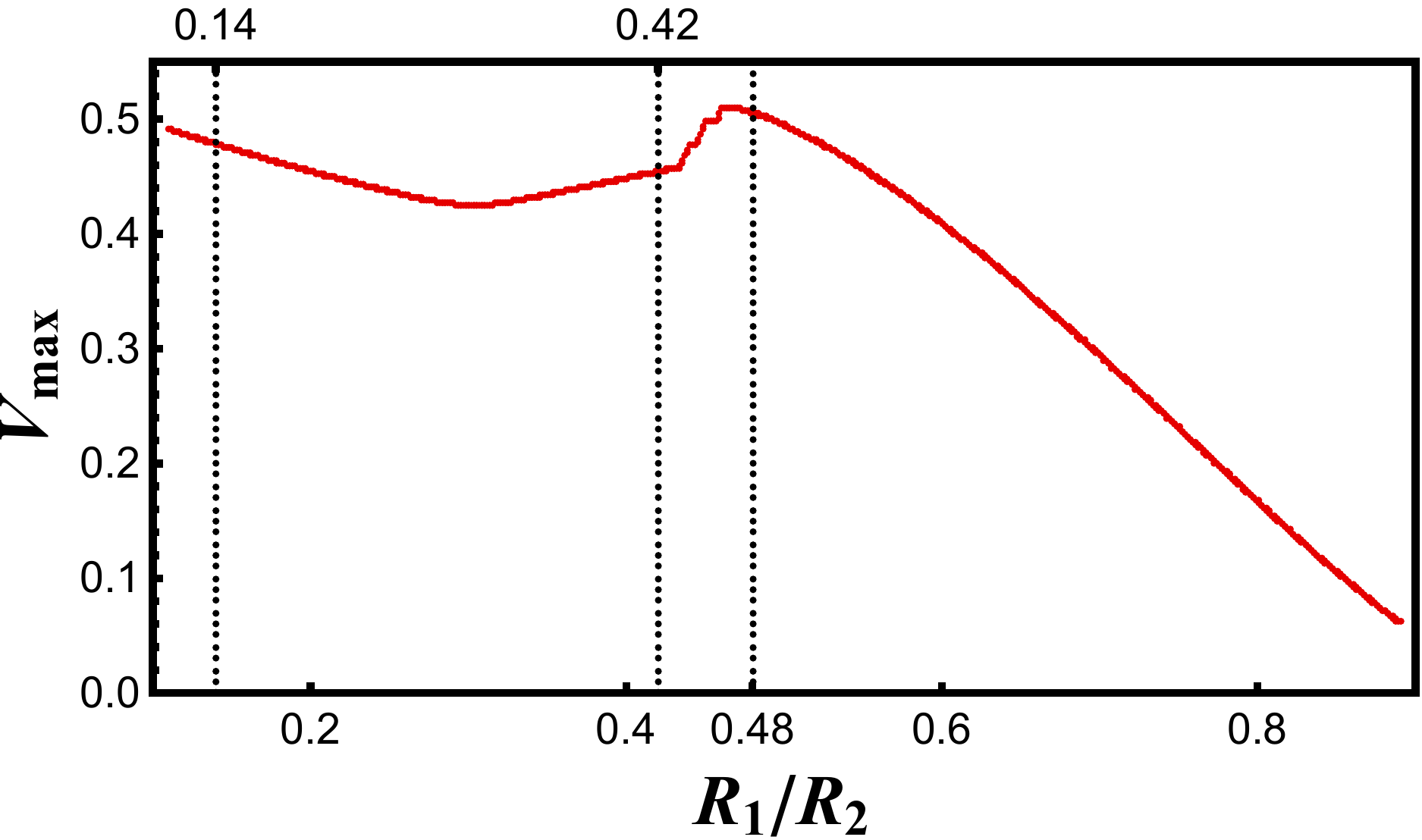}
\caption{The maximal rate of entanglement growth $V_{max}$
with different ratios of two radii $R_1/R_2$ is plotted in
red. Three dashed lines divide the whole phase diagram into three
distinct regions as discussed in Sec.\ref{sec_3}.}\label{fig_ms}
\end{figure}

\item Region with double phase transitions

For the ratio $0.42\leq R_1/R_2\leq 0.48$, it is interesting to notice that there is a peak for the borderline of two phases, therefore, in this region the HRT surface will change its phase twice when $R_1/R_2$ is fixed, as shown in Fig.\ref{fig_transitiondouble1}. At the early stage of evolution, the HRT surface is in the hemi-torus phase and at the intermediate stage, it undergoes the first phase transition from the hemi-torus phase to the two-disk phase. Eventually, the HRT surface will undergo the second phase transition to the original hemi-torus phase and will persist all the way to equilibrium.

It is also intriguing to notice that when the system approaches the first critical point (at which the phase changes from the hemi-torus phase to the two-disk phase), the time derivative of HEE seems to be continuous and this is different from the behavior of the system at the second critical point as shown in Fig.\ref{fig_transitiondouble2}. When the HRT surface undergoes the second phase transition, the time derivative of HEE is discontinuous, which is consistent with the result in the region with single phase transition. Furthermore, after passing through the second critical point, the system does not reach the equilibrium immediately. This phenomenon is in contrast to the strip case, in which the discontinuity only occurs at the equilibrium.
\end{itemize}

After identifying three distinct regions during the evolution of
HEE, we are wondering how the entanglement growth is
characterized by these regions and how the different values of the
parameters could affect these regions. Therefore, in the next two
subsections, we will explore the dependence of the rate of the
entanglement growth $V_\mathcal{A}$ on the ratio of two radii
$R_1/R_2$ as well as the dependence of the evolution of HEE on the
thickness $v_0$ of the null shell.

\subsection{Maximal rate of the entanglement growth $V_{max}$}

As discussed in \cite{Liu:2013iza,Liu:2013qca}, the rate of linear
growth at the intermediate stage provides a geometric
interpretation of the entanglement growth: during the thermal
quench, there is a wave propagating inward from the boundary
of $\mathcal{A}$. The region which has been covered by the wave is
entangled with the region outside $\mathcal{A}$, while the region
which has not been covered is generally not entangled with the
outside. Naturally, when the wave covers the whole region
$\mathcal{A}$, the saturation occurs. This phenomenon is called
``entanglement tsunami'' in literature and the speed of the
tsunami is characterized by the maximal rate of the entanglement
growth $V_{max}$ during the evolution.

Furthermore, we point out that the dependence of the tsunami speed
$V_{max}$ on the ratio $R_1/R_2$ exhibits distinct behaviors in regions with different phase transitions, as shown in
Fig.\ref{fig_ms}. In the region with no phase transition
($R_1/R_2\leq0.14 \cup R_1/R_2\geq0.48$), the tsunami speed
$V_{max}$ always decreases with the ratio $R_1/R_2$. In particular, when the ratio $R_1/R_2\rightarrow1$, the maximal rate of entanglement growth
$V_{max}$ decreases to zero. In the region with single phase
transition ($0.14\leq R_1/R_2 \leq 0.42$), the tsunami speed
$V_{max}$ decreases linearly at first and increases again after
reaching a local minimum, while in the region with double phase
transitions ($0.42\leq R_1/R_2 \leq 0.48$), the speed of
entanglement tsunami reaches a local maximum. This reveals that
the wave-front shifts with a fast rate from the boundary
$\partial\mathcal{A}$ in the region with double phase transitions.
Moreover, in a relativistic system, it is natural to expect that
the maximal rate of entanglement growth is constrained by the
causality. In this paper, the fastest rate of entanglement tsunami
occurs at $t\approx0.46$, $V_{max}=0.51$, which is smaller than the speed of light. The result is consistent with the one in \cite{Liu:2013iza,Liu:2013qca}, which exhibit a global maximal growth rate of the $4$-dimensional SAdS case under the limit of rapid quenching.

\begin{figure}
  \centering
  \includegraphics[width=250pt]{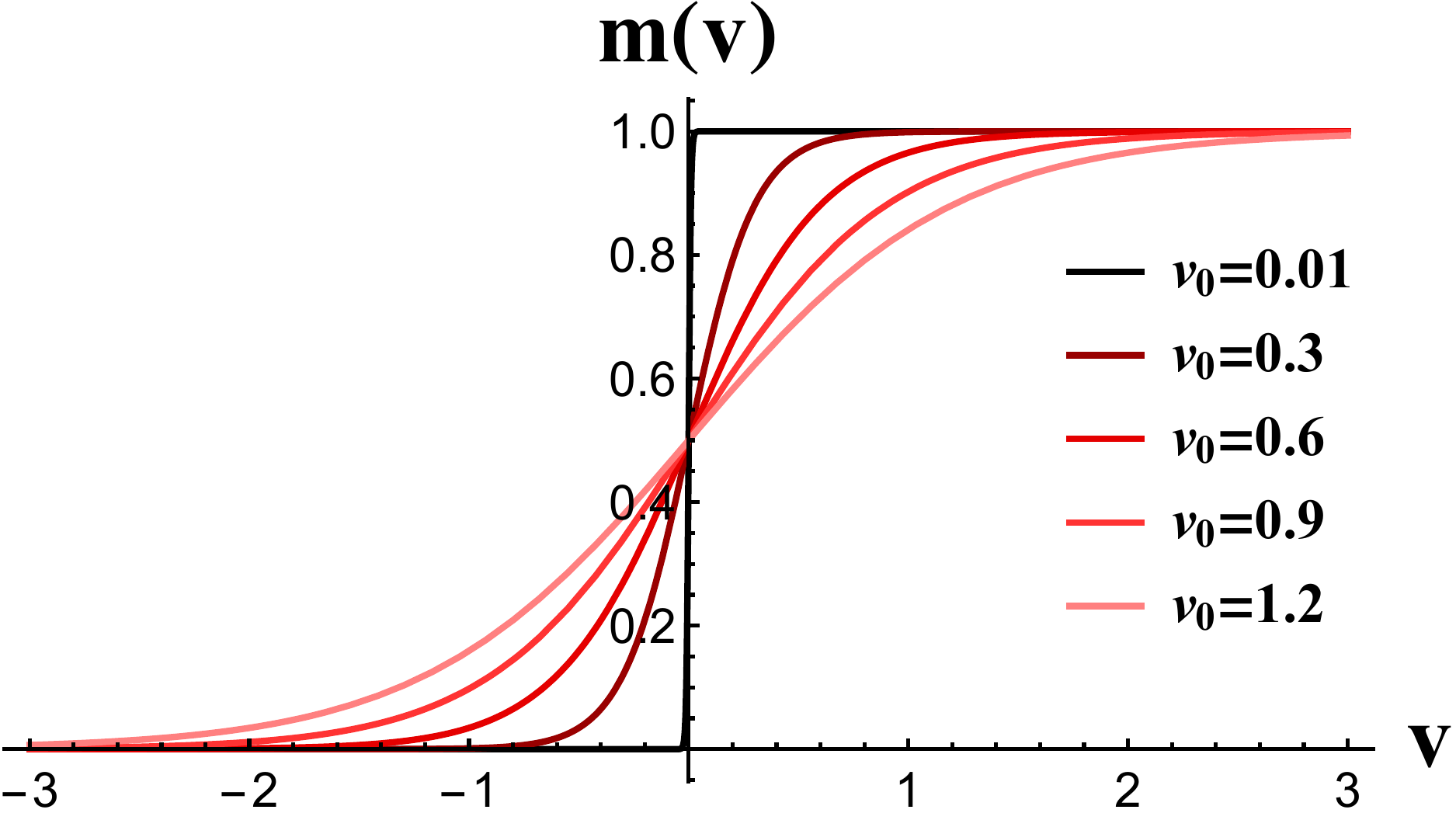}
  \caption{The evolution of the mass function $m(v)$ indicates it usually increases rapidly with small $v_0$.}\label{fig_soq}
\end{figure}

\begin{figure}
  \centering
\subfigure[]{\label{fig_t032dv}
\includegraphics[width=150pt]{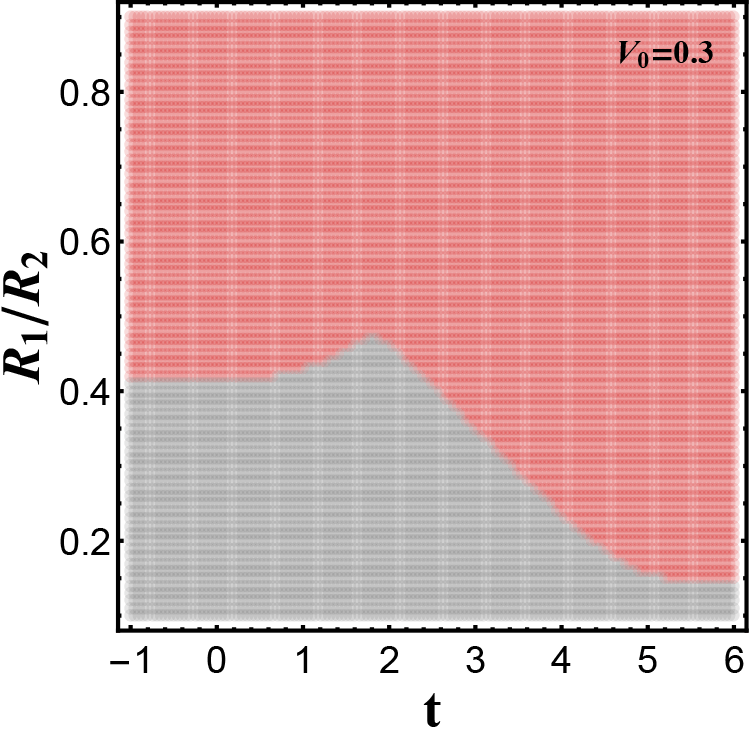}}
\hspace{0pt}
\subfigure[]{\label{fig_t062dv}
\includegraphics[width=150pt]{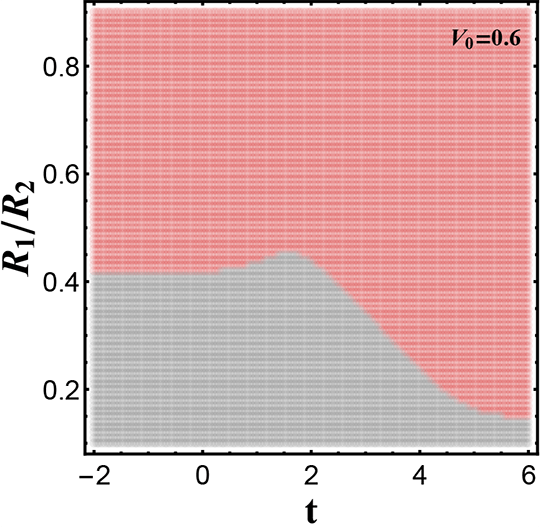}}
\hspace{0pt}
\subfigure[]{\label{fig_t122dv}
\includegraphics[width=150pt]{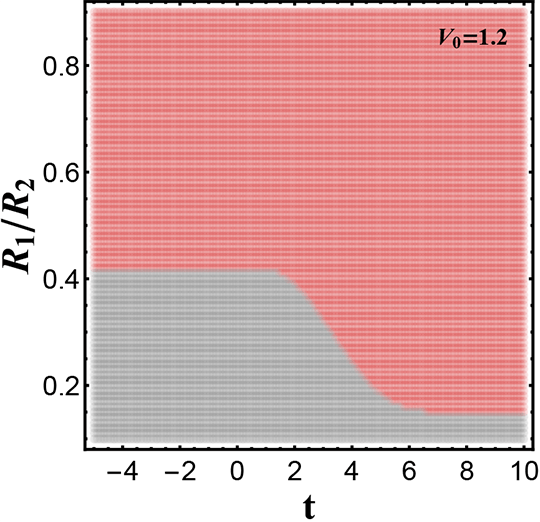}}

\caption{The evolution of HEE with various values of parameter $v_0$. The red region represents the HRT surface in the hemi-torus phase, while the grey region represents the HRT surface in the two-disk region.}\label{fig_transition2dv}
\end{figure}

\subsection{Variation of the thickness $v_0$ of the null shell}
The thickness of the null shell $v_0$ also characterizes the speed of the quench.
Defining the mass function $m(v)$ as
\begin{equation}
  m(v)\equiv\frac{M}{2}\left(1+\tanh\frac{v}{v_0}\right).
\end{equation}
For smaller $v_0$, the mass $m(v)$ of the system increases more rapidly to the final saturation, while for larger $v_0$, $m(v)$ increases slowly as shown in Fig.\ref{fig_soq}. Therefore, $1/v_0$ denotes the speed of the quench process: the larger the value of $1/v_0$ is, the sooner the quench saturates.

In Fig.\ref{fig_transition2dv}, we plot the evolution of HEE for various values of the parameter $v_0$.  The position of the critical point near equilibrium will not be affected by $v_0$. For the larger $v_0$, the subsystem $\mathcal{A}$ takes longer time to reach equilibrium.

The most prominent phenomenon in Fig.\ref{fig_transition2dv} is the size change of the region with double phase transitions. For small $v_0$, the peak of the critical point is very sharp.  With the increase of $v_0$, the peak of the critical point decreases and eventually vanishes such that the region with double phase transitions finally disappears.

\section{Conclusions and discussions}
We have investigated the holographic thermalization process of an annular subsystem $\mathcal{A}$ on the boundary over the Vaidya-AdS geometry.
Two distinct configurations of HRT surface are obtained which are the hemi-torus phase and the two-disk phase. Which phase the HRT surface belongs to depends on the ratio of the inner radius to the outer radius of the annulus.
In addition, the maximal rate of the entanglement growth $V_{max}$ exhibits distinct behavior with the different ratio of two radii.

During the thermalization process, the system with fixed $R_1/R_2$ possibly undergoes a phase transition or double phase transitions from a hemi-torus configuration to a two-disk configuration, or vice versa. The occurrence of various phase transitions is determined by the ratio of two radii of the annulus, which allows us to extract out three distinct regions. When the annulus is fairly wide or narrow, the HRT surface $\gamma_\mathcal{A}$ is always in the two-disk phase or hemi-torus phase and no phase transition occurs during the whole process, and the entanglement tsunami propagates more slowly with the larger ratio $R_1/R_2$. For the ratio $0.14\leq R_1/R_2 \leq 0.42$, the phase transition occurs one time during the thermalization, and the propagation of the entanglement tsunami reaches a local minimum.
It is quite intriguing that, there exists a region that the phase transition occurs twice during the thermalization. The HRT surface is in the hemi-torus phase at the early time, in the two-disk phase at the intermediate and in the hemi-torus phase at the late time. In this region, the propagation of the entanglement tsunami reaches a local maximum, which means the entanglement grows fairly fast in the region with double phase transitions. Moreover, the local maximum we obtained is consistent with the fastest rate of entanglement growth under the fast quenching limit, which means the rate is constrained by causality. In addition, the region with double phase transitions becomes wide with a fast quench, and becomes narrow, even vanishes with a slow quench.

In this paper, we have discussed the evolution of HEE following a global quench. It is interesting to generalize our analysis to the inhomogeneous and anisotropic case. Moreover, due to the restriction of the numeric method, we have only investigated the quench with the thickness of the null shell $v_0\geq0.3$. It is also worthy to investigate the evolution of HEE in the thin shell limit, since the diagram of entanglement tsunami is more precise than the case with a finite thickness $v_0$.

\centerline{\rule{80mm}{0.1pt}}

\end{document}